\documentclass[12pt,preprint]{aastex}

\slugcomment{Submitted to ApJ Letters }

\begin{document}

\title{Discovery of Radio Afterglow from the Most Distant Cosmic Explosion}

\author{Poonam\,Chandra\altaffilmark{1}, Dale\,A.\,Frail\altaffilmark{2}, 
 Derek\,Fox\altaffilmark{3},
 Shrinivas\,Kulkarni\altaffilmark{4}, 
Edo\,Berger\altaffilmark{5}, 
S.\,Bradley\,Cenko\altaffilmark{6}, 
 Douglas\,C.-J.\,Bock\altaffilmark{7}, 
 Fiona\,Harrsion\altaffilmark{4}, Mansi\,Kasliwal\altaffilmark{4}}

\altaffiltext{1}{Department of Physics, Royal Military College of
  Canada, Kingston, ON, Canada; Poonam.Chandra@rmc.ca}

\altaffiltext{2}{National Radio Astronomy Observatory, 1003 Lopezville
  Road, Socorro, NM 87801.}

\altaffiltext{3}{Department of Astronomy \& Astrophysics, 525 Davey
  Laboratory, Pennsylvania State University, University Park, PA
  16802.}

\altaffiltext{4}{Department of Astronomy,
  California Institute of Technology, Pasadena, CA 91125.}

\altaffiltext{5}{Harvard University,
60 Garden Street, Cambridge, MA 02138.}

\altaffiltext{6}{Department of Astronomy, 601 Campbell Hall,
  University of California, Berkeley, CA 94720-3411.}

\altaffiltext{7}{Combined Array for Research in Millimeter-wave Astronomy,
P.O. Box 968, Big Pine, CA 93513.}

\newcommand{\myemail}{Poonam.Chandra@rmc.ca}
\newcommand{\Swift}{\textit{Swift}}
\newcommand{\event}{GRB 090423}

\begin{abstract}
  We report the discovery of radio afterglow emission from the gamma-ray burst
  GRB\,090423, which exploded at a redshift of 8.3, making it the object with
  the highest known redshift in the Universe. By combining our radio
  measurements with existing X-ray and infrared observations, we
  estimate the kinetic energy of the afterglow, the geometry of the
  outflow and the density of the circumburst medium. Our best fit
  model is a quasi-spherical, high-energy explosion in a low,
  constant-density medium. \event\ had a similar energy release to the other
well-studied high redshift GRB 050904 ($z=6.26$),
but their circumburst densities differ by two orders of
magnitude. We compare the properties of \event\ with a sample of GRBs
at moderate redshifts. We find that the high energy and afterglow
properties of \event\ are not sufficiently
different from other GRBs to suggest a different kind of progenitor, such as
a Population III star.   However, we argue that 
it is not clear that the afterglow properties alone
can provide convincing identification of Population III progenitors.
We suggest that the millimeter and centimeter
radio detections of \event\ at early times contained emission from a reverse
shock component. This has important implications for the detection
of high redshift GRBs by the next generation of radio facilities.
\end{abstract}


\section{Introduction}\label{sec:intro}

Because of their extreme luminosities GRBs are detectable out to
large distances by current missions, and due to their connection to core
collapse SNe \citep{wb06}, they could in
principal reveal the stars that form from the first dark matter halos
($z\sim$ 20--30) through to the epoch of reionization at $z=11\pm 3$ and
closer \citep{lr00,cl00b,gmaz04,ioc07}.   
As bright continuum sources,
GRB afterglows  also make  ideal backlights to probe the intergalactic medium
as well as the interstellar medium in their host galaxies.   Predicted to
occur at redshifts beyond those where quasars are expected, they could
be used to study both the reionization history and metal
enrichment of the early universe \citep{tkk+06}.

The fraction of detectable GRBs that lie at high
 redshift ($z>6$) is, however, expected to be small
($<$10\%; \citealt{pcb+09,bl06}). Until recently there were only two
GRBs with measured redshifts $z>6$;  GRB\,050904 \citep{kka+06} and GRB\,080913
\citep{gkf+09} with $z=6.3$ and $z=6.7$, respectively. However, on
April 23, 2009 the \Swift\ Burst Alert Telescope (BAT) discovered \event\ 
and the on-board X-ray Telescope (XRT) detected and localized a
variable X-ray afterglow \citep{tfl+09,sdc+09}. In ground-based
follow-up observations no optical counterpart was found but a fading
afterglow was detected by several groups at wavelengths longward of J
band (1.2 $\mu$m). Based on both broadband photometry and near
infrared (NIR) spectroscopy, the sharp optical/NIR drop off was argued
to be due to the Lyman-$\alpha$ absorption in the intergalactic
medium, consistent with a redshift with a best-fit value of
$z=8.26^{+0.07}_{-0.08}$ \citep{tfl+09}.  The high redshift of \event\ 
makes it the most distant observed GRB, as well as the most distant
object of any kind other than the Cosmic Microwave Background.  This event occurred 
approximately 630 million years after the Big Bang, confirming that massive stellar
formation occurred in the very early universe.

In this paper we report the discovery of the radio afterglow from \event\
with the Very Large Array\footnote{The Very Large Array is
operated by the National Radio Astronomy Observatory, a facility of
the National Science Foundation operated under cooperative agreement
by Associated Universities, Inc.} (VLA).   Broadband
afterglow observations provide constraints on the explosion energetics,
geometry, and  immediate environs of the progenitor star. The afterglow 
has a predictable temporal and spectral
evolution that depends on the kinetic energy and geometry of the shock,
the density structure of the circumburst environment, and 
shock microphysical parameters which depend on the physics of particle
acceleration and the circumburst magnetic field. To the degree that we
can predict differences in the explosion and circumburst media between
GRB progenitors at high and low redshifts, we can search for these
different signatures in their afterglows. This has been the motivation
for previous multi-wavelength modeling of the highest-$z$
afterglows \citep{fck+06,gfm07}.

In order to investigate the nature of the \event\ explosion,
we combine our radio measurements with published X-ray and 
NIR observations, and apply a model of the blast wave  evolution to
fit the afterglow data (\S\ref{sec:results}).   We compare the explosion energetics,
circumburst density, and other derived characteristics to a sample
of well-studied events, and discuss prospects for using afterglow measurements
to investigate the nature of high-$z$ massive star progenitors.   

\section{Observations}\label{sec:obs}

\subsection{Radio observations}\label{sec:radio}

We began observing a field centered at the NIR afterglow of \event\ 
with the VLA about one day after the burst \citep{cfk09}. Our first
detection of the GRB afterglow
was not until about one week later at a flux density of
$73.8\pm21.7$ $\mu$Jy.  We continued to monitor the GRB with the VLA
until it faded below detection on day 64. 
The flux density scale was tied to
3C\,286 and the phase was measured by switching with a 6.5 minute
cycle time between the GRB field and the point source calibrator
J0954+177. To maximize sensitivity, the full VLA continuum bandwidth
(100 MHz) was recorded in two 50 MHz bands.  Data reduction was
carried out following standard practice in the {\it AIPS} software
package.

In Table \ref{tab:radio} we list the {\it individual} VLA
measurements.  In order to improve our detection sensitivity, we
averaged several adjacent observations. 
Datasets were combined in the
{\it UV}-plane prior to imaging. By averaging  three adjacent
epochs (2009 May 1--May 3) when the afterglow was brightest, we estimate
the best GRB position by fitting a 2-D Gaussian, which is; RA, Dec
(J2000): $09^h55^m(33.279\pm0.005)^s,\;18^d08'(57.935\pm0.067)''$.
This position is consistent with an earlier, less accurate WFCAM-UKIRT
position from \citet{tlk+09}.

Table \ref{tab:radio-comb} gives the flux densities at the averaged epochs. 
For all epochs the flux density
was measured at the position given above. We plot these data in
Fig.~\ref{fig:multi}.  There is a broad plateau of about 45 $\mu$Jy
from 12 to 38 days, followed by a decline around day 55. The initial
detections on days 8--10 could have contribution from a short-lived
reverse shock (\S\ref{sec:results}).

We also observed \event\ with the Combined Array for Research in
Millimeter-wave Astronomy (CARMA) at 95 GHz band on 2009 Apr 25.19 UT.
The observation was 8 hours in length. Data was obtained under
non-ideal weather conditions. 
The peak flux at the VLA afterglow position 
is 450$\pm$ 180 $\mu$Jy. \citet{cbw+09} reported a secure
millimeter band detection ($\lambda=3$ mm) at a flux
density of $200$ $\mu$Jy with the Plateau de Bure Interferometer
(PdBI) observed on 2009 Apr 23 \& 24. \citet{rwb+09} placed flux upper
limit of 0.96 mJy in 250 MHz band observed
with the Max-Planck-Millimeter Bolometer (MAMBO-2) array at the IRAM
30-m telescope.  The Westerbork Synthesis Radio Telescope (WSRT) also
observed the GRB between 2009 May 22.48 UT to 23.46 UT at 4.9 GHz and did
not detect the afterglow \citep{van09}.

\subsection{X-ray observations}\label{sec:xray}

{\it Swift}-XRT \citep{bhn+05} 
observed
the field of \event\ for one week in Photon Counting (PC) mode.  The
XRT light curve is obtained from the on-line
repository\footnote{http://www.Swift.ac.uk/xrt\_curves}
\citep{ebp+07}.  The X-ray spectrum is well-fit by a power-law model
with a photon index $\Gamma = 2.05^{+0.14}_{-0.09}$ and a total column
density of $N_H=(8.7\pm2.5) \times 10^{20}$ cm$^{-2}$
\citep{tfl+09,kpp+09}. We converted the $0.3-10.0$ keV counts to a
flux density at $E = 1.5$\,keV ($\nu_{0} = 3.6 \times 10^{17}$\,Hz)
using the above value for $\Gamma$ and an unabsorbed count rate conversion
of 1 count $=4.6\times 10^{-11}$ erg/cm$^2$/s.

\subsection{Near Infrared Observations}\label{sec:nir}

The NIR  afterglow was observed by a variety of facilities
worldwide; we have used values from \citet{tfl+09}.  To convert
magnitudes to flux densities, we used zero-point measurements from
\citet{fsi95}. 
We have incorporated the Galactic extinction ($E(B-V) =
0.029$; \citealt{dl90}) into these results.

\section{Results and Afterglow Modelling}
\label{sec:results}

Here we combine our radio data with the existing X-ray and NIR data and model
the afterglow evolution, interpreting it in terms of the relativistic
blast wave model \citep{mes06}.  In this model the afterglow physics
is governed by the isotropic kinetic energy of the blast wave shock
$E_{K, \rm iso}$, the jet opening angle $\theta_j$, the density of the
circumburst medium $n$, and the microscopic parameters such as
electron energy index $p$, and the fraction of the shock energy
density in relativistic electrons $\epsilon_e$ and magnetic fields
$\epsilon_B$.  The afterglow modeling software \citep{yhsf03} assumes
a standard synchrotron forward shock formulation.  In the X-ray band,
we exclude the data before $\sim$ 3900 s, since it contains a flare
which is more likely due to the GRB itself than the afterglow.

\begin{figure}
\includegraphics[angle=0,width=0.84\textwidth]{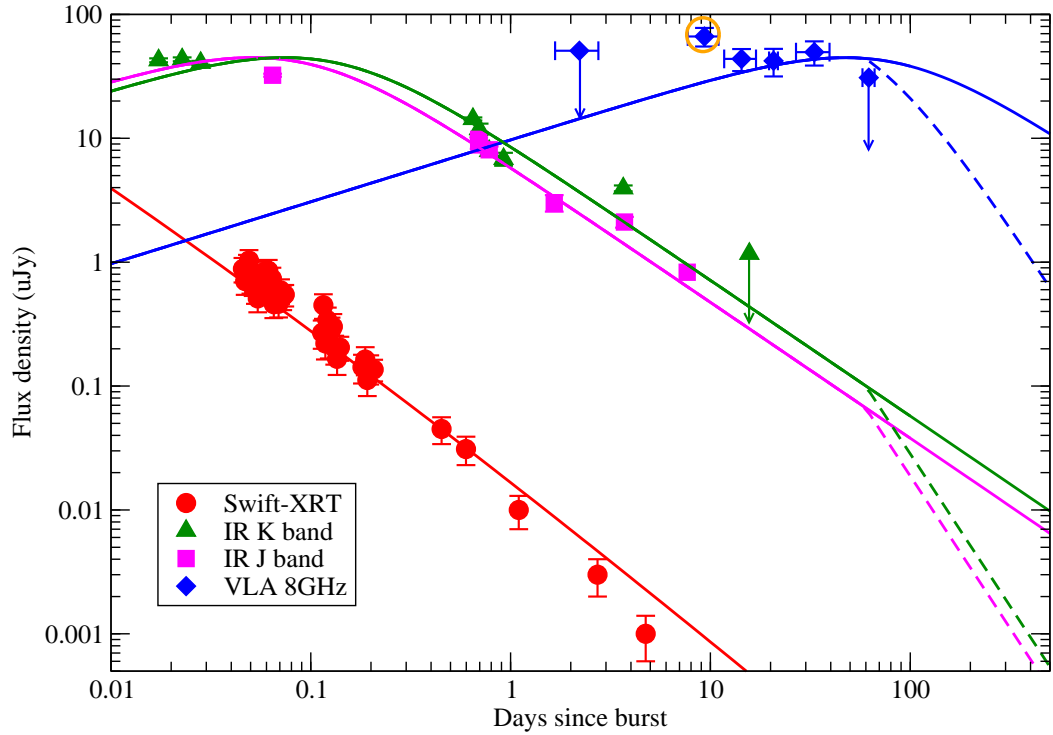}
\caption{Multiwaveband observations for GRB\,090423. The solid lines are 
  best fit light curves for constant density isotropic model. The
  orange circled radio data likely has a contribution from RS
  (\S \ref{sec:results}).  
Dashed lines show model with a possible jet break around $t_j=45$ d.}
\label{fig:multi}
\end{figure}

It is well known \citep{spn98,sph99,cl99} that the afterglow framework
allows the above blast wave parameters to be constrained using multi-wavelength light
curves.  First, we note the constancy of the peak flux density ($F_{\nu, \rm
  max}$) between the NIR and the radio bands in
Fig.~\ref{fig:multi}. If we interpret this as the passage of the
synchrotron peak frequency $\nu_m$ through each band, this immediately
rules out the wind model ($F_{\nu,\rm max} \propto t^{-1/2}$) and
favors a constant-density ambient medium ($F_{\nu,\rm max} \propto
t^0$).  Another related constraint which comes from the
Fig.~\ref{fig:multi} is the time of the peak in the NIR versus the
radio bands. We note that in the NIR band, the light curve peaks at
$\sim 0.08$ d.  Thus if there was an early jet break the model
predicts that the synchrotron peak frequency $\nu_m$ would evolve from
NIR to radio band around day 10 ($\nu_m \propto t^{-2}$), however, for
the isotropic model $\nu_m$ should pass through the 8.5 GHz band
around $\sim 50 $ days ($\nu_m \propto t^{-3/2}$). Since radio light
curve indeed peaks at about 50 d, this confirms that the jet break has not
occurred at least  until the afterglow peaked in radio band.

Second, the declining part of the IR light curve is well fit by a
power law with a decay index $\alpha=-1.10\pm0.27$.  Whereas the
overall X-ray light curve after 3900s is well fit with a power law
index of $\alpha=-1.35\pm0.15$. For the isotropic, constant density
model we expect the flux at a given frequency $\nu_{\rm obs}$ to
decline as $t^{3(1-p)/4}$ for $\nu_m<\nu_{\rm obs}<\nu_c$ and
$t^{(2-3p)/4}$ for $\nu_m<\nu_c<\nu_{\rm obs}$, where $\nu_c$ is the
synchrotron cooling frequency 
\citep{spn98}. These relations give consistent values of $p$ for the
NIR ($p=2.46\pm0.36$) and X-ray ($p=2.46\pm0.20$).

Finally, having obtained an estimation for $p$ and determining the
fact the X-ray frequency has evolved past the cooling frequency $\nu_{\rm X-ray} >
\nu_c$, we can put a constraint on the total energy carried by the
fireball electrons ($\epsilon_e E$) just from a single X-ray flux
measurement. Using Eq. 4 of \citet{fw01} and X-ray flux on day 1,
we obtain the fireball electron energy per unit solid angle in an opening
angle   $1/\Gamma$ on $t=1$ d to be  
$\epsilon_e E
/4 \pi = 7.4 \times 10^{51}$ ergs. If we assume $\epsilon_e=1/3$ \citep{fw01},
then the total fireball energy per unit solid angle in this opening 
will be $E/4 \pi = 2.5 \times 10^{52}$ ergs.

We summarize our robust inferences based on this preliminary analysis:
(a) the data favors an isotropic explosion in a constant density
medium; (b) the cooling frequency lies between the IR and X-ray bands;
(c) the afterglow kinetic energy is large. From \citet{tfl+09} we also
know that the extinction due to a putative host galaxy is negligible
($A_V<0.08$).

We now move on to more detailed modeling \citep{yhsf03} guided
broadly by these preliminary results. We fit a constant density model
for parameters: $E_{K, \rm iso}$, $\theta_j$, $n$, $p$, $\epsilon_e$
and $\epsilon_B$.  All parameters were allowed to vary freely except
that we fixed $p=2.46$ to lie in a narrow range ($\pm0.20$). The best
fit parameters are tabulated in Table \ref{tab:param}. Our best fit
model is plotted in Fig.~\ref{fig:multi}. 

This simple model provides a reasonable fit to the data. The model
implies GRB kinetic energy to be $E_{K}=3.8\times 10^{53}$ erg.  
However, the last measured data point is around day 65 (radio band) and
the last detections in the radio and NIR bands are at about day 40 and day
46, respectively. Therefore a late jet break cannot be ruled out by
these data. To illustrate this more concretely we overlay our best-fit
model in Fig.~\ref{fig:multi} with a late jet break $t_j\sim 45$ d.
The implied jet opening angle $\theta_j> 0.21$ rad reduces both the
radiated and the kinetic energies of this event by a factor of $\sim
45$. In this case the isotropic equivalent gamma-ray energy
E$_{\gamma}=1\times{10^{53}}$ erg \citep{von09} and the blastwave
kinetic energy $E_{K}=3.8^{+9.8}_{-1.7}\times 10^{53}$ erg give {\it
  lower limits} to the beaming-corrected values of
$E_\gamma>2.2\times{10^{51}}$ erg and $E_K>8.4^{+21.6}_{-3.7}\times
10^{51}$ erg, respectively.

 The radio data point on day 9.34 ($t \sim 1$ d in the rest frame)
has high flux and does
not go through the best fit forward shock model. 
Such early, short-lived radio emission is
fairly common in GRBs at lower redshifts
and is thought to be due to a contribution from afterglow reverse shock (RS)
\citep{kfs+99,sr03,np04}.
We can make a
rough estimate of the peak RS contribution for \event\
using the formulation of \citet{np04} and the  best fit
parameters. The order of magnitude calculation shows that the
RS  contribution is expected to be $\sim
20 $ $\mu$Jy. Even though this estimate may have large uncertainties,
this does support our speculation that the RS
is likely contributed some of the radio emission seen on $t=9.34$ d.
If we assume that
the data on day 9.34 represents the peak of the reverse shock then this
corresponds to  a synchrotron self absorption frequency of
$\nu_a^r\sim 3.4 \times 10^{14}$
Hz. Using the scaling law for RS emission,
 $t_{\rm radio}= \nu_a^r t_o/\nu_{\rm radio}$,
the time for RS peak in 90 GHz is $t \sim 0.87$ d.
This implies that PdBI
data flux reported by \citet{cbw+09} 
 may also have contribution from the RS.

\section{Discussion and Conclusions}\label{sec:dis}

\event\ is the highest-redshift object for which we have
multi-wavelength observations, including good quality radio measurements.
 Below we address the following questions:
based on its afterglow properties what can we
learn about properties of the explosion and environs for this
highest-redshift GRB? And, can we identify any differences between high
and low redshift GRBs which indicate that they might arise from different progenitors?
In particular, the initial generations of stars in the early universe are thought to
be brighter, hotter and more massive ($>100\, M_\odot$) than stars
today \citep{hai08,byhm09}.   Detecting these so-called
Population III (Pop III) stars is one of the central observational challenges in
modern cosmology,  and the best prospect appears to be through observing
their stellar death \citep{hfw+03} via a supernovae (SNe) or 
gamma-ray burst explosion.   It is worth asking what observational
signatures could signal a Pop III GRB.

\begin{figure}
  \includegraphics[angle=0,width=0.88\textwidth]{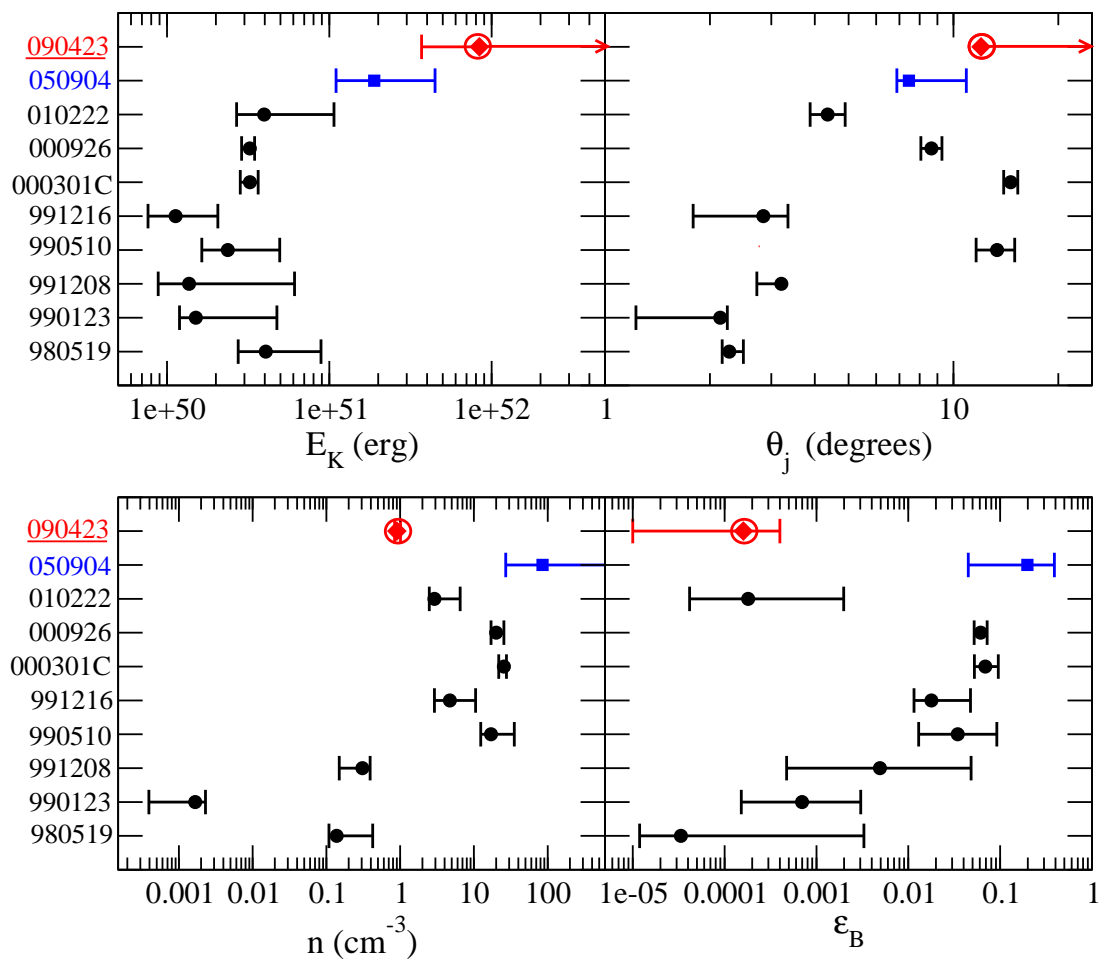}
\caption{
  Comparison of \event\ best fit parameters 
  with few moderate-$z$ GRBs ($z\sim 1-3 $) 
  from \citet{pk01} and with the high-$z$ GRB 050904 ($z=6.295$,
   \citealt{gfm07,fck+06}). Here the upper limit on \event\ 
$E_K$ is $(3.8+9.8)\times10^{51}$ erg.}
\label{fig:comparison}
\end{figure}

Other than \event ,  only one other $z > 6$ event, GRB\,050904 ($z=6.26$), 
has high quality broadband afterglow measurements.
In Fig.~\ref{fig:comparison} we plot the best-fit parameters of these
two GRBs along with a sample of well-studied lower redshift events from \citet{pk01}. 
Both high redshift bursts stand out in terms of their large blast wave energy
($> 10^{52}$ erg). We know from samples of well-studied afterglows
\citep{fks+01,pk01,yhsf03}, that most  have radiative and kinetic
energies of order $\sim 10^{51}$ erg.  In the collapsar model the jet
kinetic energy from a Pop III GRB could be 10--100 times larger than a Population II (Pop  II) 
event  \citep{fwh01,hfw+03}.   However, an energetic
explosion does not appear to be an exclusive property of high-$z$ 
GRBs. There is a small but growing population of bursts with energy 
$> 10^{52}$ erg, termed 'hyper-energetic GRBs'
\citep{cfh+09}, which includes moderate-$z$ events like GRB\,070125
\citep{ccf+08}  and GRB\,050820A \citep{ckh+06}.

Another potentially useful diagnostic is the density structure in the
immediate environs of the progenitor star. The radio data is a
sensitive {\it in situ} probe of the density because its emission
samples the optically thick part of the synchrotron spectrum. The
afterglows of \event\ and GRB\,050904 are best fit by a constant
density medium and not one that is shaped by stellar mass loss
\citep{cl99}.  However, many afterglows at all redshifts
are best fit by a constant density medium (e.g.~\citealt{yhsf03}).
The density obtained for GRB\,050904 was the highest seen ($n\approx84-680$
cm$^{-3}$) for any GRB to date, while \event\ with $n=0.9$ cm$^{-3}$
does not stand out (Fig.~\ref{fig:comparison}), indicating these two 
high redshift bursts exploded in very different environments. A circumburst density
of order unity is predicted for Pop III stars, since this density is  limited
by strong radiation pressure in the mini halo from which the star 
was formed \citep{byh03}.   This is not an unique property,
since many local SNe explode in tenuous media, and so density constraints 
are not useful to signal Pop III explosions.

For the other afterglow parameters ($p$, $\epsilon_e$, $\epsilon_B$
and $\theta_j$) there are no published predictions for how they may
differ between different progenitor models. Thus we turn to considering
the prompt high-energy emission of \event.

\citet{sdc+09} and \citet{tfl+09} both noted that the high-energy
properties of \event\ (fluence, luminosity, duration, 
radiative energy) are not substantially different from those
moderate-$z$ GRBs.  We are not aware of any
quantitative predictions based on these observed properties that would
discriminate between Pop\,II and Pop\,III progenitors. For example,
apart from the effect of ($1+z$) time dilation, there is no reason to
expect high-$z$ GRBs to have significantly longer intrinsic durations.
Collapsar models which form black holes promptly through accretion
onto the proto-neutron star (Type I) or via direct
massive ($>260$ M$_\odot$) black hole formation (Type III) have
durations set by jet propagation and disk viscosity timescales,
respectively \citep{mwh01,fwh01},
which are $\sim 10$ s in the rest frame (with large uncertainties;
\citealt{fwh01}).

Metallicity can also be an important discriminant. There is a critical
metallicity ($Z>10^{-3.5}\,Z_\odot$) below which high-mass Pop III
stars dominate \citep{byhm09,bl06}.  The contribution of Pop III stars
to the co-moving star formation rate is expected to peak around $z=15$
but their redshift distribution exhibits a considerable spread to
$z\sim$7. Thus we might find high-redshift GRBs with Pop III
progenitors in ``pockets'' of low metallicity. 
\citet{sdc+09} argue for a lower bound of $Z>0.04\,Z_\odot$ based on
their detection of excess soft X-ray absorption by metals along the
line of sight, in comparison to the Milky Way column density predicted
from \ion{H}{1} (21\,cm) measurements.  
We do not
consider this a robust measurement as it is sensitive to a range of
unaccounted-for systematic effects, including: spectral variability;
spectral curvature; low-amplitude X-ray flares; and the presence of
intervening (cosmological) absorption systems along the line of sight.

Summarizing the above discussion, we do not find that the individual
properties of \event\ are sufficiently dissimilar to other GRBs to
warrant identifying it as anything other than a normal GRB. We lack
robust predictions of well-defined afterglow signatures that could
allow us to unambiguously identify a Pop III progenitor star from its
afterglow properties alone. Significantly larger numbers of GRBs at high
redshift with  well-sampled afterglow light curves,
high-resolution spectra, and host galaxy detections are needed to
determine if high redshift GRB progenitors differ in a statistical sense from 
those at low redshift.

We note that, like GRB\,050904, the \event\ afterglow indicates 
the signature of  reverse shock (RS)
emission in the radio, as seen in the VLA  and  PdBI data.  
\citep{ioc07} have studied the
expected RS emission at high redshift, and they find that the effects 
of time dilation almost
compensate for frequency redshift , resulting in a near-constant 
observed peak frequency in the mm band
($\nu \sim 200$~GHz) at a few hours post-event, and a
 flux at this frequency that is almost independent of redshift.
Further, the mm band does not suffer significantly either 
from extinction (in contrast to the optical)
or scintillation (in contrast to the radio).
Therefore, detection of mm flux at a few hours post event
 should  be a good method of indicating a high
redshift explosion.  ALMA, with its high sensitivity
 ($\sim$75 $\mu$Jy in 4 min), will be a potential tool for 
selecting potential high-$z$ bursts that would be 
high priority for intense followup across the spectrum.
This will hopefully greatly increase the rate at which high-$z$ 
events are identified.

Finally, our data does not rule out a late jet break at $t_j>45$ d, which, as discussed 
above, makes the total explosion energy uncertain. Extremely sensitive
VLA observations would be required to distinguish between the isotropic versus jet
model. However, in 2010 with an order of magnitude enhanced sensitivity the EVLA will
be the perfect instrument for such studies. For a  2\,hr integration
in 8 GHz band, the EVLA can
reach sensitivity up to 2.3 $\mu$Jy which will be able to detect the 
\event\ for 2 years or 6 months if the burst is isotropic or jet-like, 
respectively.
EVLA will thus be able detect fainter events and
  follow events like GRB\,050904 and \event\ for a longer duration, therefore
obtaining  better density measurements, better estimates of outflow geometry  
 and the total kinetic  energy.
 
\acknowledgments

We thank Bob Dickman for his generous allocation of VLA time and Joan
Wrobel and Mark Claussen for the timely scheduling of the
observations. This work made use of data supplied by the UK Swift
Science Data Centre at the University of Leicester.

\newpage

\begin{deluxetable}{lcccclc}
\tablecaption{Radio observations of \event\ 
\label{tab:radio}}
\tablewidth{0pt}
\tablehead{
\colhead{Date} & \colhead{$\Delta t$} & \colhead{Tel.} & \colhead{Freq.} & \colhead{$F_\nu$\tablenotemark{a}} & \colhead{Int.} & \colhead{Array}\\
\colhead{UT} & \colhead{days} & \colhead{} & \colhead{GHz} & \colhead{$\mu$Jy} & \colhead{time\tablenotemark{b}} & \colhead{Conf.}
}
\startdata
Apr 25.01 & 1.68  & VLA& 8.46 &    $51\pm 45$ &$13$ & B\\
Apr 26.08 & 2.75  & VLA& 8.46 &    $-17 \pm 37$ &16 & B\\
May 01.05 & 7.72  & VLA& 8.46 &   $74  \pm   22$  &$50 $  & B      \\
May 03.08 & 9.75 & VLA & 8.46 &   $77   \pm  18$  & $75$& B\\
May 03.98 & 10.65 & VLA & 8.46 & $57 \pm 19$ & $75$ & B\\
May 05.05 & 11.72 & VLA & 8.46 & $38\pm  19$  &$71$ & B\\
May 05.99 & 12.66 & VLA & 8.46 & $87 \pm 23$ &$56$ & B \\
May 08.09 & 14.76 & VLA& 8.46 &    $-4 \pm    19$ & $67$      & B  \\
May 09.05 & 15.72 & VLA& 8.46 &    $5   \pm  18$   &  $71$   & B  \\
May 10.08 & 16.75 & VLA& 8.46 &   $73    \pm  18$   & $71$ & B     \\
May 12.99 & 19.66 & VLA & 8.46 &   $29 \pm 18$ & $70$ & B\\
May 14.10 & 20.77 & VLA & 8.46 &  $88\pm  21$ & $45$ & B\\
May 15.05 & 21.72 & VLA & 8.46 &   $7 \pm 15$ & $110$ & B\\
May 20.13 & 26.80 & VLA & 8.46 &  $42 \pm 18$ & $78$ & B\\
May 27.12 & 33.79 & VLA & 8.46 &  $78 \pm 19$ & $77$ & BnC\\
Jun 01.11 & 38.78 & VLA & 8.46 &  $44\pm 18$ & $77$ & BnC\\
Jun 20.00 & 57.67 & VLA & 8.46 & $19\pm 21$ & $78$ & C\\
Jun 26.08 & 63.75 & VLA & 8.46 & $49\pm 19$ & $78$ & C\\
Jun 26.91 & 64.58 & VLA & 8.46 & $-4\pm 20$ & $75$ & C\\
\hline
Apr 25.20 & 1.87  & CARMA & 92.5 & $450 \pm 180$ &\ldots & \ldots \\
\enddata
\tablenotetext{a}{Peak flux density at \event\ position.}
\tablenotetext{b}{Integration time on \event\ in minutes.}
\end{deluxetable}

\newpage

\begin{deluxetable}{lll}
\tablecaption{VLA 8.5 GHz flux densities of \event\
at combined epochs
\label{tab:radio-comb}}
\tablewidth{0pt}
\tablehead{
\colhead{Epochs} & \colhead{Days since} &  \colhead{Flux
density}\\
\colhead{Combined} & \colhead{explosion} &  \colhead{$\mu$Jy}
}
\startdata
Apr 25.01--Apr 26.08 & $2.21\pm0.54$  &  $50.9\pm30.9$  \\
May 01.05--May 03.98 & $9.34\pm1.64$  &  $66.4\pm11.4$ \\
May 05.05--May 10.08 & $14.32\pm2.60$ &  $43.7\pm8.9$  \\
May 12.99--May 15.05 & $20.71\pm1.06$ &  $42.2\pm10.6$ \\
May 20.13--Jun 01.11 & $33.12\pm6.32$ &  $49.6\pm11.0$ \\
Jun 20.00--Jun 26.91 & $62.00\pm4.33$ &  $7.8\pm11.6$  \\
\enddata
\end{deluxetable}

\newpage

\begin{deluxetable}{lll}
\tablecaption{Best fit parameters for multiwaveband modeling of \event\ for $p=2.46\pm0.2$
\label{tab:param}}
\tablewidth{0pt}
\tablehead{
\colhead{Parameters}&\colhead{Isotropic}&{Jet ($t_j>45$ d)}
}
\startdata
$E_{\gamma}$ (ergs) & $1.0 \times 10^{53}$ & $>2.2 \times 10^{51}$\\
$E_{K}$ (ergs) & $3.8^{+9.8}_{-1.7}\times 10^{53}$ & $>8.4^{+21.6}_{-3.7}\times
10^{51}$\\
$n$ (cm$^{-3}$) & $0.90^{+0.11}_{-0.06}$ & \ldots\\
$\epsilon_B(\%)$ & $0.016^{+0.024}_{-0.015}$& \ldots\\
$\epsilon_e$ & $0.28^{+0.10}_{-0.01}$ & \ldots\\
\enddata
\tablecomments{For jet model ($t_j>45$ d), 
the $n$, $\epsilon_B$ and $\epsilon_e$ were fixed to
 the best fit 
isotropic model values. }
\end{deluxetable}

\newpage

\newpage

\end{document}